\documentclass[showpacs,amsmath,amssymb,aps,twocolumn]{revtex4-1}
\usepackage{graphics}
\usepackage{graphicx}
\usepackage{txfonts}
\usepackage{bm}   
\usepackage{amsmath,mathrsfs}
\usepackage{multirow}
\usepackage{tablefootnote}

\begin{document}

\title{A novel nuclear dependence of nucleon-nucleon short-range correlations}

\author{ Hongkai Dai$^{1,2}$ }
\author{ Rong Wang$^{2,3,4}$ }\email{Corresponding author: rwang@impcas.ac.cn}
\author{ Yin Huang$^3$ }
\author{ Xurong Chen$^{2}$ }

\affiliation{
$^1$ College of Physics and Electronic Engineering, Northwest Normal University,
Lanzhou 730070, China\\
$^2$ Institute of Modern Physics, Chinese Academy of Sciences, Lanzhou 730000, China\\
$^3$ Lanzhou University, Lanzhou 730000, China\\
$^4$ University of Chinese Academy of Sciences, Beijing 100049, China\\
}

\date{\today}

\begin{abstract}
A linear correlation is found between the magnitude of nucleon-nucleon
short-range correlations and the nuclear binding energy per nucleon with
pairing energy removed. By using this relation, the strengths of
nucleon-nucleon short-range correlations of some unmeasured nuclei are predicted.
Discussions on nucleon-nucleon pairing energy and
nucleon-nucleon short-range correlations are made.
The found nuclear dependence of nucleon-nucleon short-range correlations
may shed some lights on the short-range structure of nucleus.
\end{abstract}

\pacs{25.30.Fj, 25.40.Ep, 21.30.-x}

\maketitle

\section{Introduction}
\label{Sec:Intro}

In the nuclear shell model, nucleons undergo nearly independent motion in the mean field
created by their mutually attractive potential and the residual interactions can be
treated by perturbation theory. However, the hard repulsive core of nucleon-nucleon
(N-N) interactions generate nucleons with momentum above the Fermi momentum when nucleons
move very close together. These short-distance structures are commonly called short-range
correlations (SRC). The prominent feature of N-N SRC is that the correlated
pair has large relative momentum and small center-of-mass (c.m.) momentum. There is no
doubt that N-N SRC is a major source of the high-momentum tail for the nucleon momentum
distribution in a nucleus. Moreover a nucleon in a two-nucleon SRC pair reaches close
to the repulsive core of the N-N potential, which creates a very high local
density. A complete understanding of SRC is one of the fundamental goals
of nuclear physics and would help to complete our knowledge about the complex structure
of nuclei and to probe cold dense nuclear matter\cite{jlab-science,SRC-Nstar}.

Details of SRC structure are beginning to be unveiled through decades of
dedicated theoretical\cite{PLB1977,PR1981,NPA1987,PRC1992,ARNPS2001,PRC1996,Alvioli2016,
Schiavilla2007,Sargsian2005,Vanhalst2011,Vanhalst2012,Colle2015} and
experimental\cite{jlab-science,PRC1993,PLB1999,BNL,CLAS,Piasetzky2006,hallC,Science2014,JLabRev,ExpRev}
efforts. Two types of significant experiments are the two-nucleon
knockout process\cite{jlab-science,PLB1999,BNL,Piasetzky2006,Science2014}
and inclusive electron scattering in the large $x\sim 2$ region\cite{PRC1993,CLAS,hallC}.
When the N-N SRC pair is knocked out, the transverse momenta of the two nucleons of the pair
are in the opposite directions with high relative transverse momenta,
leaving the remnant nucleus nearly undisturbed.
Although suffering from very low count rates, exclusive pair knock out experiments have
advantages in studying c.m. motion of SRC pair, the relative momenta of the nucleons,
and the isospin structure of the pair.
The electron inclusive cross section in the scaling region of $1.5<x<2$ also
detects the SRC pairs by interacting with the high momentum nucleons.
Although it can not distinguish types of the SRC pairs,
the cross section in the scaling region reflects the relative strength
of N-N SRC in a nucleus. So far the inclusive cross section ratios of heavy nuclei
to light nuclei have been measured at SLAC and JLab\cite{PRC1993,hallC}.

The quantitative measurements of the inclusive cross section in the scaling\cite{PRC1993}
(independence of $x$ and $Q^2$) region $1.5<x<2$ of different nuclei
provide a good opportunity to study the nuclear dependence of SRC.
The common scale quantifying the strength of N-N SRC was thought to be
the average nuclear density. However, the average nuclear density fails
to predict the strength of N-N SRC with accuracy, because the shapes
of nuclei are quite different and sometimes greatly deformed from
the sphere shape. In theory, N-N SRC is related to the nucleon spectral function
$S(\vec{k},E)$ at high momentum. The local density
approximation\cite{Antonov1986,Benhar1994,Sick1994,Benhar1989,Benhar2008}
is used extensively to evaluate the nucleon spectral function.
It very successfully reproduces and predicts the cross section of
inclusive quasi-elastic electron-nucleus scattering.
An early SLAC experiment indicated that the EMC effect is
proportional to the average nuclear density \cite{emc-slac}.
However, it has been shown that the local environment is the most relevant quantity
for explaining the EMC effect by recent measurement at JLab\cite{Seely2009}
and by the detailed analysis\cite{NuclDepenRev}.
Recently, the nuclear binding energy and the strong
interaction energy among nucleons are used to study the nuclear dependence
of the EMC effect\cite{Wang2015}. Since there is a good
linear correlation found between N-N SRC and the EMC effect, it is worthwhile
to find the relations between the N-N SRC and the nuclear binding energy.

Study of the nuclear dependence of N-N SRC is of significance, which
could unveil the underlying mechanism of N-N SRC.
Though the N-N SRC probabilities inside various nuclei are hard to calculate
so far, the first step is to find a quantity with which the N-N SRC scales.
Nuclear binding energy is an important static property of a nucleus,
which comes from the average effect of the complicated and different interactions
among nucleons. How do the nucleon-nucleon attractions contribute
to the formation of the nucleon in the high momentum tail of the distribution?
What are the primary interactions which contribute to the N-N SRC?
These are the goals of the study. In this paper, we try to find a scaling relation
which can describes the N-N SRC of various nuclei.

The paper is organized as follows. In Sec. \ref{Sec:SRCs},
we review some theoretical investigations and the experimental measurements
of N-N SRC. In Sec. \ref{Sec:Binding}, we introduce a modified nuclear
binding energy $B^{\text{Mod}}$ with the nuclear pairing effect removed.
In Sec. \ref{Sec:NuclDepen}, a novel linear correlation is shown between
the strength of N-N SRC and the modified nuclear binding energy per nucleon.
Finally, discussions and a simple summary are given in Sec. \ref{Sec:Discu}.

\section{Nucleon-nucleon short-range correlations}
\label{Sec:SRCs}

The unambiguous evidence of N-N SRC comes from the nucleon knock-out process
in high energy scattering experiments with high momentum
transfer and high missing momentum\cite{jlab-science,PLB1999,BNL,Piasetzky2006,Science2014}.
The high missing momentum is
predominantly balanced by a single recoiling nucleon, which shows
strong back-to-back directional correlation of nucleon-nucleon pairs.
The two-nucleon knockout process is not only useful to identify N-N SRC
but also important to study the isospin structure of SRC pairs.
The measurement of the two nucleon knockout reaction by the high energy
electron probe reveals that proton-neutron pairs are almost 20 times
as prevalent as proton-proton pairs in $^{12}$C\cite{jlab-science}.
Recently it is reported that p-n SRC pairs are dominant in all measured nuclei,
including the neutron-rich imbalanced nuclei such as lead and iron\cite{Science2014}.
It is also shown that the high
momentum (greater than Fermi momentum) of the nucleon is attributed to the correlated
pairs rather than the Pauli exclusion principle, which explains the
higher average proton kinetic energy in neutron-rich nuclei\cite{Sargsian2014}.

The primary goal of N-N SRC experiments is now to map out the strength
of the correlated dense structure in nuclei and its isospin structure.
The inclusive measurement of quasielastic electron scattering\cite{PRC1993,CLAS,hallC}
in the kinematic region
of $x>1$ and $Q^2>1$ GeV$^2$ is a good way to study the strength of N-N SRC.
L. L. Frankfurt et al. point out that the $x$-scaling relation in the region
of $1.4<x<2$ is directly connected to the N-N SRC\cite{PRC1993}.
Assuming that the nucleon-nucleon correlations dominate in large $x$ region ($x>1$),
the inclusive cross section of quasielastic scattering can be decomposed as
\begin{equation}
\sigma^A(x,Q^2)=\sum_{j=2}^A A\frac{1}{j}a_j(A)\sigma_j(x,Q^2),\\
\label{InclusvieFor}
\end{equation}
where $\sigma_j(x,Q^2)$ is the cross section of electron scattering on the
$j$-nucleons correlated and compact configuration ( $\sigma_j(x>j,Q^2)=0$ ),
and $a_j(A)$ is proportional to the probability
of finding a nucleon in a $j$-nucleons correlation.
By the assumption that the probability of $j$-nucleons SRC drops rapidly with $j$,
the $x$- and $Q^2$-scaling of $\sigma^A/\sigma^D$ is expected in region of $1< x \le 2$.
Therefore the SRC ratio $a_2=2\sigma^A/(A\sigma^D)$ in the scaling region
is used to quantify the relative strength of N-N SRC in nucleus. The $a_2$ values
used in this work are measured as the ratios of per-nucleon inclusive electron
scattering cross sections of nuclei to that of deuteron in the region
of large $x$ ($1.5<x<2$).
To quantify the strength of N-N SRC more accurately in experiment,
$R_{2N}$ is given by removing the smearing effect of the c.m. motion
of the correlated pair\cite{hallC}. The combined data of $a_2$ and $R_{2N}$
measured at SLAC and JLab \cite{NuclDepenRev}
are listed in Table~\ref{tab:Data}.

\begin{table}[htp]
\centering
\caption{
The per-nucleon nuclear binding energies and the measured SRC ratios
of various nuclei are shown.
The binding energies are taken from Ref.~\cite{atomic-mass}
and the combined data of $a_2$ and $R_{2N}$ are taken from Ref.~\cite{NuclDepenRev}.
}
\begin{tabular}{ccccc}
\hline
nucleus & $B/A$ (MeV) & $B^{\text{Mod}}/A$ (MeV) & $a_{2}$  & $R_{2N}$ \\
\hline
$^{3}$He    & 2.573   &  2.573  &  2.13$\pm$0.04  & 1.92$\pm$0.09 \\
$^{4}$He    & 7.074   &  5.574  &  3.57$\pm$0.09  & 2.94$\pm$0.14 \\
$^{9}$Be    & 6.463   &  6.463  &  3.91$\pm$0.12  & 3.37$\pm$0.17 \\
$^{12}$C    & 7.680   &  7.391  &  4.65$\pm$0.14  & 3.89$\pm$0.18 \\
$^{27}$Al   & 8.332   &  8.332  &  5.30$\pm$0.60  & 4.40$\pm$0.60 \\
$^{56}$Fe   & 8.790   &  8.761  &  4.75$\pm$0.29  & 3.97$\pm$0.34 \\
Cu, \footnote{
For the copper target, both $^{63}$Cu and $^{65}$Cu have big natural abundances.
The weighted mean binding energy and weighted modified binding energy
are used for Cu target.}       & 8.754   &  8.754  &  5.21$\pm$0.20  & 4.33$\pm$0.28 \\
$^{197}$Au  & 7.916   &  7.916  &  5.13$\pm$0.21  & 4.21$\pm$0.26 \\
\hline
\end{tabular}
\label{tab:Data}
\end{table}

On the theoretical side, the N-N SRC in the nucleus is thought to be mainly governed by
the N-N tensor force\cite{PRC1992,ARNPS2001,Schiavilla2007,PRC1996,Alvioli2016,
Sargsian2005,Vanhalst2011,Vanhalst2012,Colle2015}.
The high-momentum feature and the strength of N-N SRC are studied using the variational
Monte Carlo method\cite{PRC1992,ARNPS2001,Schiavilla2007},
the nucleon spectral function approach\cite{PRC1996,Alvioli2016},
the decay function formalism for two-nucleon emission reactions\cite{Sargsian2005},
and counting correlated pairs residing in a relative-$S$ state
with many-body correlation operator implemented\cite{Vanhalst2011,Vanhalst2012,Colle2015}.
High nucleon momentum and high local density are two basic characteristics of N-N SRC pairing.
The contribution of N-N SRC to the high-momentum nucleons is described by $R_{2N}$.
If $R_{2N}$ is only sensitive to spin 1 pairs and the correlated pairs
are mainly in $S$ state, then $R_{2N}(N_{pn}+N_{pp}+N_{nn})/N_{pn(^3S_1)}$
is a good variable for describing high local density configurations,
where $^3S_1$ denotes that the correlated pair is in a spin-triplet state.
The number of correlated pairs $N_{pn}$, $N_{pp}$, $N_{nn}$, and $N_{pn(^3S_1)}$
can be estimated by counting pairs in $S_{n,l=0}$
state\cite{Vanhalst2011,Vanhalst2012,Colle2015,CosynDiscu}.

A linear correlation between N-N SRC and the EMC slope
is found\cite{NuclDepenRev,src-emc1,src-emc2,src-emc3,src-emc4},
thanks to the recent measurements of very light nuclei at JLab.
The EMC effect reflects the nuclear modifications of valence quark distributions
inside nucleons. The N-N overlapping configuration surely plays an important
role in the measured SRC ratios and the EMC effect.
The found linear correlation indicates that both effects are sensitive
to the high local density or the high virtuality (momentum) of the nucleon.
However for heavy nuclei, the EMC-SRC correlation shows
some non-linearity\cite{src-emc2,src-emc3,src-emc4}.
The $a_2$ of $^{197}$Au is slightly smaller than that of $^{56}$Fe, but
the EMC slope of $^{197}$Au is obviously larger than that of $^{56}$Fe.

\section{Nucleon-nucleon interaction and binding energy}
\label{Sec:Binding}

Detailed study of nuclear force is a fundamental and interesting subject
in nuclear physics, which accounts for the properties and dynamics of nuclei.
Although it is acknowledged that the N-N interaction comes from
the residual strong force and the underlying theory is Quantum Chromodynamics (QCD),
resolving the N-N interaction from quark-gluon freedoms is still difficult because of
confinement and asymptotic freedom in QCD.
Nevertheless, there are models which can be used to describe the nuclear force effectively.
The long-range attractive interaction is well studied by the $\pi$ exchange process
predicted by Yukawa, and the intermediate interaction is governed by $2\pi$, $\rho$,
$\omega$ and $\sigma$ exchanges. The detailed form of the N-N potential
is now given by a phenomenological potential from experiments\cite{Reid93,AV18,Born},
Lattice QCD calculations\cite{LQCD1,LQCD2},
and the Chiral Perturbation Theory (ChPT)\cite{ChPT-rev1,ChPT-rev2,NN-ChPT-1,NN-ChPT-2}.
However, in terms of the intermediate attraction and the repulsive core at
short distance, there still exist big uncertainties between different models.

A simple and straightforward description of the strength of the N-N interaction
in the nucleus is the nuclear binding energy. Although the binding energy is
a simple static quantity for the nucleus, it actually depicts the average strength
of the nuclear force and the compactness among nucleons. The nuclear binding energy
is defined as $B=ZM(^1H)+NM(^1n)-M(A,Z)$, which is related to the atomic mass
directly. The atomic masses are precisely measured world wide
and evaluated\cite{atomic-mass}. In Ref. \cite{atomic-mass2}, we can find
the complete information of experimental data evaluated (used and rejected).
For the world average data, the least-squares adjustment procedure is used to
determine the best values for the atomic masses and their uncertainties\cite{atomic-mass2}.

A nucleus is a perfect interacting Fermi liquid, and a simple estimate of
nuclear binding energy is the semi-empirical Bethe-Weizs\"{a}cker (BW)
mass formula\cite{BW-ref}. According to the BW formula, the binding energy of
a nucleus of atomic mass number $A$ and proton number $Z$ is described as
\begin{equation}
B(A,Z)=a_vA-a_sA^{2/3}-a_cZ(Z-1)A^{-1/3}-a_{sym}(A-2Z)^2A^{-1}+\delta,
\label{BWMF}
\end{equation}
where $a_v=15.79$ MeV, $a_s=18.34$ MeV, $a_c=0.71$ MeV, $a_{sym}=23.21$ MeV,
and $\delta$ is the pairing energy correction.
In order to study the short-distance structure of nucleus,
we define a modified nuclear binding energy $B^{\text{Mod}}$ as
\begin{equation}
B^{\text{Mod}}(A,Z)=B(A,Z)-\delta,\\
\label{ModBDef}
\end{equation}
where $B(A,Z)$ is the precisely measured nuclear binding energy\cite{atomic-mass}
and $\delta$ is the pairing energy in the BW formula\cite{BW-ref}.
$\delta$ is given by,
\begin{equation}
\delta=\left\{
\begin{array}{cc}
a_{p}A^{-1/2}  & \text{(even Z, even N)}\\
0              & \text{(odd A)}       \\
-a_{p}A^{-1/2} & \text{(odd Z, odd N)} \\
\end{array}\right.   \\
\label{deltaDef}
\end{equation}
in which $a_p=12$ MeV.
The per-nucleon nuclear binding energies and the modified binding energies
of the targets are listed in Table \ref{tab:Data}.

The nuclear pairing effect is one of the long-standing problems of nuclear structure,
which was first investigated in the even-odd staggering of binding energy
decades ago\cite{pair-rev-1,pair-rev-2,pair-rev-3,pair-rev-4,pair-theo-beyond}.
It is not a surprise that conditions for pairing are satisfied in nuclei,
for the nuclear interaction between identical nucleons is attractive in $S$-channel.
In analogy to the Cooper pairs in the microscopic theory of superconductivity,
the nucleon pairing can be explained in the Bardeen-Cooper-Schrieffer (BCS) approximation
with interactions between nucleons in conjugate states\cite{pair-rev-1,BCS}.
However, the nuclear system is different from the superconducting metal in the finite
size of the fermi gas system. The present advanced models are the
Hartree-Fock-Bogoliubov (HFB) approximation\cite{HFB-1,HFB-2},
and the Lipkin-Nogami (LN) method\cite{LN-L,LN-N,LN-2}, which are
all based on a self-consistent mean-field representation.
The nuclear pairing arising from the mean-field effect may be
not important for the N-N SRC, since short-range correlation
is a rare phenomenon beyond the mean-field description.

\section{Nuclear dependence of N-N SRC}
\label{Sec:NuclDepen}

\begin{figure}[htp]
\begin{center}
\includegraphics[width=0.45\textwidth]{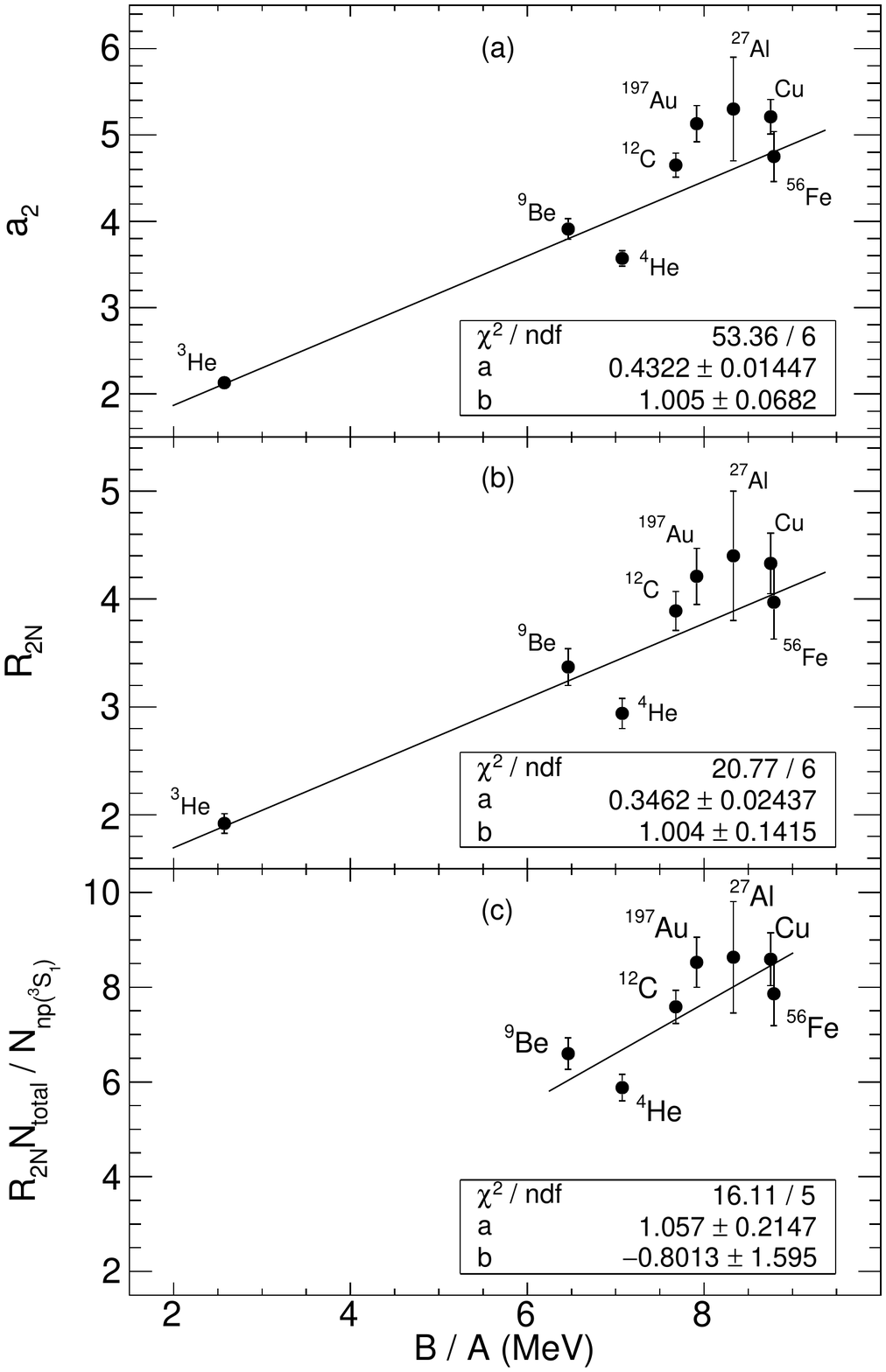}
\caption{
(a) $a_2$ as a function of binding energy per nucleon;
(b) $R_{2N}$ as a function of binding energy per nucleon;
(c) $R_{2N}N_{total}/N_{np(^3S_1)}$ as a function of binding energy per nucleon.
}
\label{fig:SRC_B}
\end{center}
\end{figure}

\begin{figure}[htp]
\begin{center}
\includegraphics[width=0.45\textwidth]{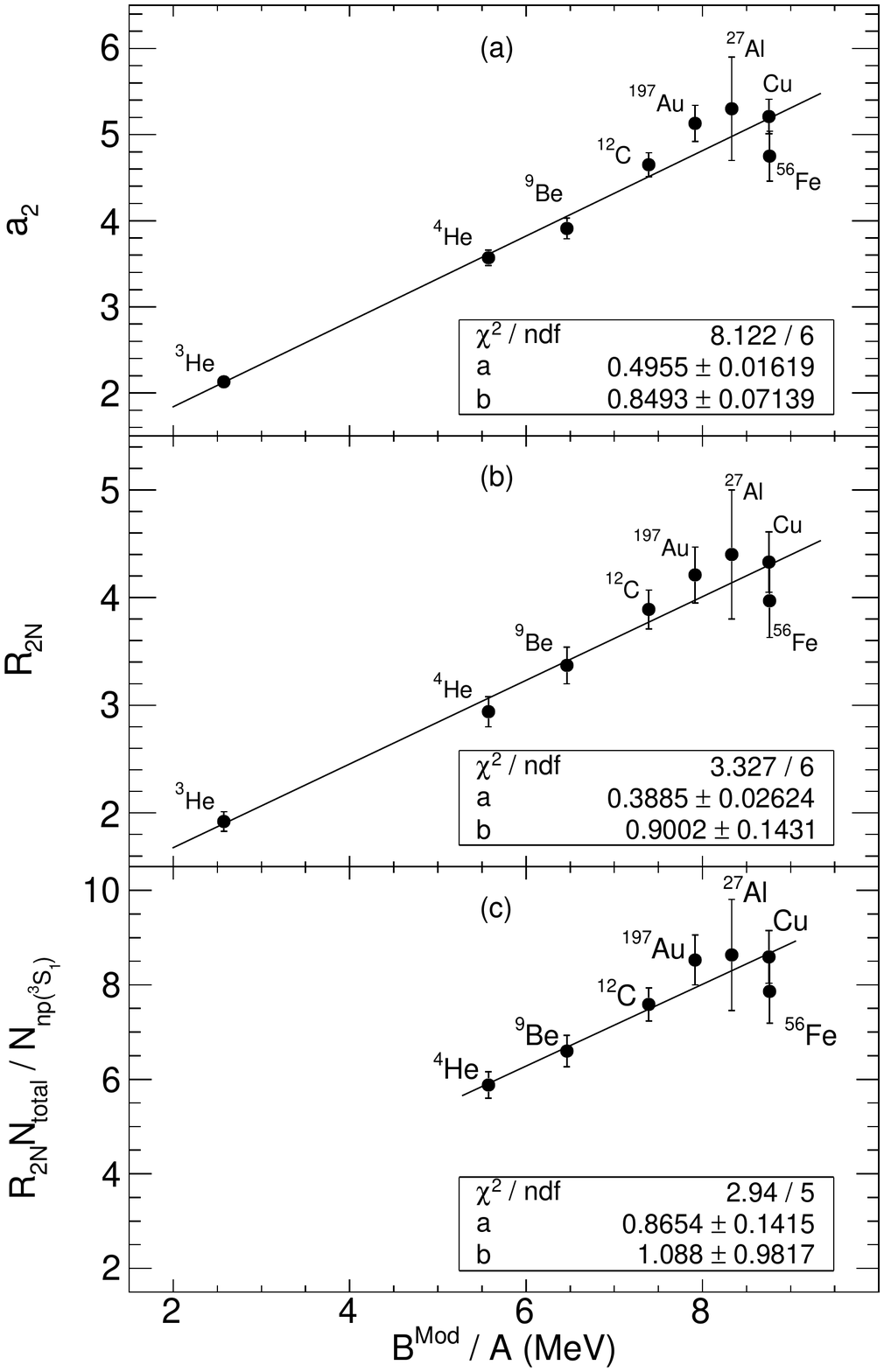}
\caption{
(a) $a_2$ as a function of modified binding energy per nucleon;
(b) $R_{2N}$ as a function of modified binding energy per nucleon;
(c) $R_{2N}N_{total}/N_{np(^3S_1)}$ as a function of modified binding energy per nucleon.
}
\label{fig:SRC_ModB}
\end{center}
\end{figure}

Fig. \ref{fig:SRC_B} shows the N-N SRC ratios as functions of the nuclear
binding energy per nucleon. Roughly, the strength of N-N SRC relates with
the per-nucleon binding energy. But there is no clear correlation
between SRC and nuclear binding. Linear fits are performed to N-N SRC
ratios versus the per-nucleon binding energy. The qualities of the fits are poor
with $\chi^2/N_{df}=$ 8.9, 3.5 and 3.2 for $a_2$ vs. $B$,
$R_{2N}$ vs. $B$ and $R_{2N}N_{total}/N_{np(^3S_1)}$ vs. $B$ respectively.
$^4$He and $^9$Be are at odds with the linear correlation
between N-N SRC and nuclear binding. The per-nucleon binding energy of $^4$He
is much larger than that of $^9$Be, however the measured N-N SRC ratio of $^4$He
is obviously smaller than that of $^9$Be. The general nuclear binding energy
is not a good scale to describe the phenomenon of N-N SRC.

Fig. \ref{fig:SRC_ModB} shows the correlations between the N-N SRC ratios and
the modified nuclear binding energy per nucleon which has the pairing energy
subtracted. To our surprise a startling linear correlation is found between
the N-N SRC ratios and the modified nuclear binding $B^{\text{Mod}}$.
Linear fits are performed to these correlations using the formula as $a(B^{\text{Mod}}/A)+b$.
The obtained $\chi^2/N_{df}$ are 1.4, 0.55 and 0.59 for
linear fits to $a_2$ vs. $B^{\text{Mod}}$, $R_{2N}$ vs. $B^{\text{Mod}}$
and $R_{2N}N_{total}/N_{np(^3S_1)}$ vs. $B^{\text{Mod}}$ respectively.
The fitted parameters for the linear correlations
are shown in Fig. \ref{fig:SRC_ModB}. The $R_{2N}N_{total}/N_{np(^3S_1)}$ value
of $^3$He is not given in the figures, for the theoretical
estimation of $N_{total}$ and $N_{np(^3S_1)}$ for $^3$He is not realized so far.
Compared to the EMC slopes, the uncertainties of modified nuclear binding
energies are not included in the fits, which makes the observed
linear correlation more significant.

The found linear correlations suggest that the modified nuclear binding
energy per nucleon in Eq.~(\ref{ModBDef}) is a good scale to quantify the strength of
N-N SRC, although the underlying mechanism is not clear. Therefore
a new nuclear dependence of N-N SRC can be written as,
\begin{equation}
\begin{aligned}
a_2=(0.496\pm 0.017)\frac{B^{\text{Mod}}}{A~\text{MeV}}+(0.849\pm 0.072),\\
R_{2N}=(0.389\pm 0.027)\frac{B^{\text{Mod}}}{A~\text{MeV}}+(0.90\pm 0.15).\\
\end{aligned}
\label{CorrelFormula}
\end{equation}
Here we do not give the nuclear dependence of $R_{2N}N_{total}/N_{np(^3S_1)}$,
because $R_{2N}N_{total}/N_{np(^3S_1)}$ is related to the theoretical calculations
of the counting numbers of different types of N-N SRC pairs which is not
a pure experimental quantity.

\section{Discussions and summary}
\label{Sec:Discu}

A linear correlation between N-N SRC and the modified nuclear binding
with the pairing effect removed is found. The linear correlation gives a novel
nuclear dependence of N-N SRC, which can be used to predict the SRC
ratios of various nuclei. In the near future, new experiments of N-N SRC
on light, medium and heavy nuclei will be performed at JLab\cite{PR12-11-112,PR12-06-105}.
These measurements will provide a good opportunity to test the found correlation.
The predictions of SRC ratios of some nuclei are listed in Table \ref{tab:Prediction}.
The SRC ratio of $^3$H is shown in Table \ref{tab:Prediction},
which is quite close to that of $^3$He.
Up to now, the N-N SRC have been obtained only for odd-$A$ nuclei or nuclei of
even-$Z$ and even-$N$. To further study the modified nuclear binding dependence
of N-N SRC, we suggest experimental measurements of nuclei of odd-$N$ and odd-$Z$,
such as $^{6}$Li, $^{10}$B and $^{14}$N. The predicted SRC ratios
of $^{6}$Li, $^{10}$B and $^{14}$N are listed in Table \ref{tab:Prediction}
using Eq.~(\ref{CorrelFormula}).

\begin{table}[htp]
\centering
\caption{
Predictions of SRC ratios $a_2$ and $R_{2N}$ of various nuclei (some
of them never measured) by the linear correlation
between N-N SRC and modified nuclear binding energy.
}
\begin{tabular}{ccccc}
\hline
nucleus & $B/A$ (MeV) & $B^{\text{Mod}}/A$ (MeV) & $a_{2}$  & $R_{2N}$ \\
\hline
$^{3}$H   & 2.827    &  2.827  &  2.25$\pm$0.12  & 2.00$\pm$0.23 \\
$^{6}$Li  & 5.332    &  6.148  &  3.90$\pm$0.18  & 3.29$\pm$0.32 \\
$^{7}$Li  & 5.606    &  5.606  &  3.63$\pm$0.17  & 3.08$\pm$0.30 \\
$^{9}$Be  & 6.463    &  6.463  &  4.05$\pm$0.18  & 3.41$\pm$0.32 \\
$^{10}$B  & 6.475    &  6.854  &  4.25$\pm$0.19  & 3.57$\pm$0.34 \\
$^{11}$B  & 6.928    &  6.928  &  4.29$\pm$0.19  & 3.59$\pm$0.34 \\
$^{12}$C  & 7.680    &  7.391  &  4.51$\pm$0.20  & 3.78$\pm$0.35 \\
$^{14}$N  & 7.476    &  7.705  &  4.67$\pm$0.21  & 3.90$\pm$0.36 \\
$^{40}$Ca  & 8.551    &  8.504  &  5.07$\pm$0.22  & 4.21$\pm$0.38 \\
$^{48}$Ca  & 8.667    &  8.631  &  5.13$\pm$0.22  & 4.26$\pm$0.38 \\
$^{63}$Cu  & 8.752    &  8.752  &  5.19$\pm$0.22  & 4.30$\pm$0.39 \\
\hline
\end{tabular}
\label{tab:Prediction}
\end{table}

Based on the linear correlation between N-N SRC and the modified nuclear
binding, it is a reasonable conclusion that the pairing effect is the least
relevant mechanism for the observed N-N SRC. Actually, the nuclear pairing is
very different from the nucleon-nucleon short-range correlation for two reasons.
Firstly, nuclear pairing is mainly between p-p and n-n pairs, but ninety percent of
SRC correlated pairs are p-n pairs\cite{jlab-science,Science2014}. This is also
suggested by theoretical calculations applying tensor interactions of N-N system
\cite{Schiavilla2007,Sargsian2005,Vanhalst2011,Vanhalst2012}.
Secondly, N-N SRC directly relates to the
nucleon-nucleon overlapping wave functions, however, the nuclear pairing is
a mean-field effect. The correlation length of nuclear pairing is
comparable to the size of nucleus. On the one hand, the N-N SRC involves
with the significantly stronger interaction than the nuclear pairing energy.
On the other hand, the nuclear pairing is not related to the repulsive potential
at short distance. Eq.~(\ref{deltaDef}) is based on the assumption that
the pairing energies are the same for both p-p and n-n pairs.
Detailed calculation of pairing energy will give more insights about
the difference between the N-N SRC and the nuclear pairing.

The N-N SRC is a noticeable nuclear phenomenon beyond the mean field description
of nucleons. To understand the underlying mechanism of formation of
short-range correlated pairs, we are interested to find out the particular
and direct interactions which attribute to it. Inspired by recent work\cite{Wang2015},
both the Coulomb contribution and the pairing energy are removed from the nuclear
binding energy to construct a new modified binding energy $(B-\delta-B^{\text{Coul}})$.
However the correlation between SRC and $(B-\delta-B^{\text{Coul}})$ shows no
improvements. Therefore the Coulomb interaction affects the formation of N-N SRC.
The N-N SRC is connected to the high virtuality and the high local density.
No doubt that the binding energy describes the average virtuality for nucleons.
If high virtuality plays an important role in N-N SRC, then the average virtuality
should correlate with SRC. We speculate that the modified nuclear binding energy
$B^{\text{Mod}}$ is sensitive to the average local density. The larger $B^{\text{Mod}}$
becomes, the higher probability of the nucleon overcoming the hard core repulsion,
and so closer the nucleons come together. A dynamical balance between the repulsive
core and the intermediate attraction is speculated to be important in the formation
of N-N SRC pairs. Removing the pairing correction from the binding energy allows
the nuclear binding to better reflects the dynamics of N-N SRC.

\begin{figure}[htp]
\begin{center}
\includegraphics[width=0.45\textwidth]{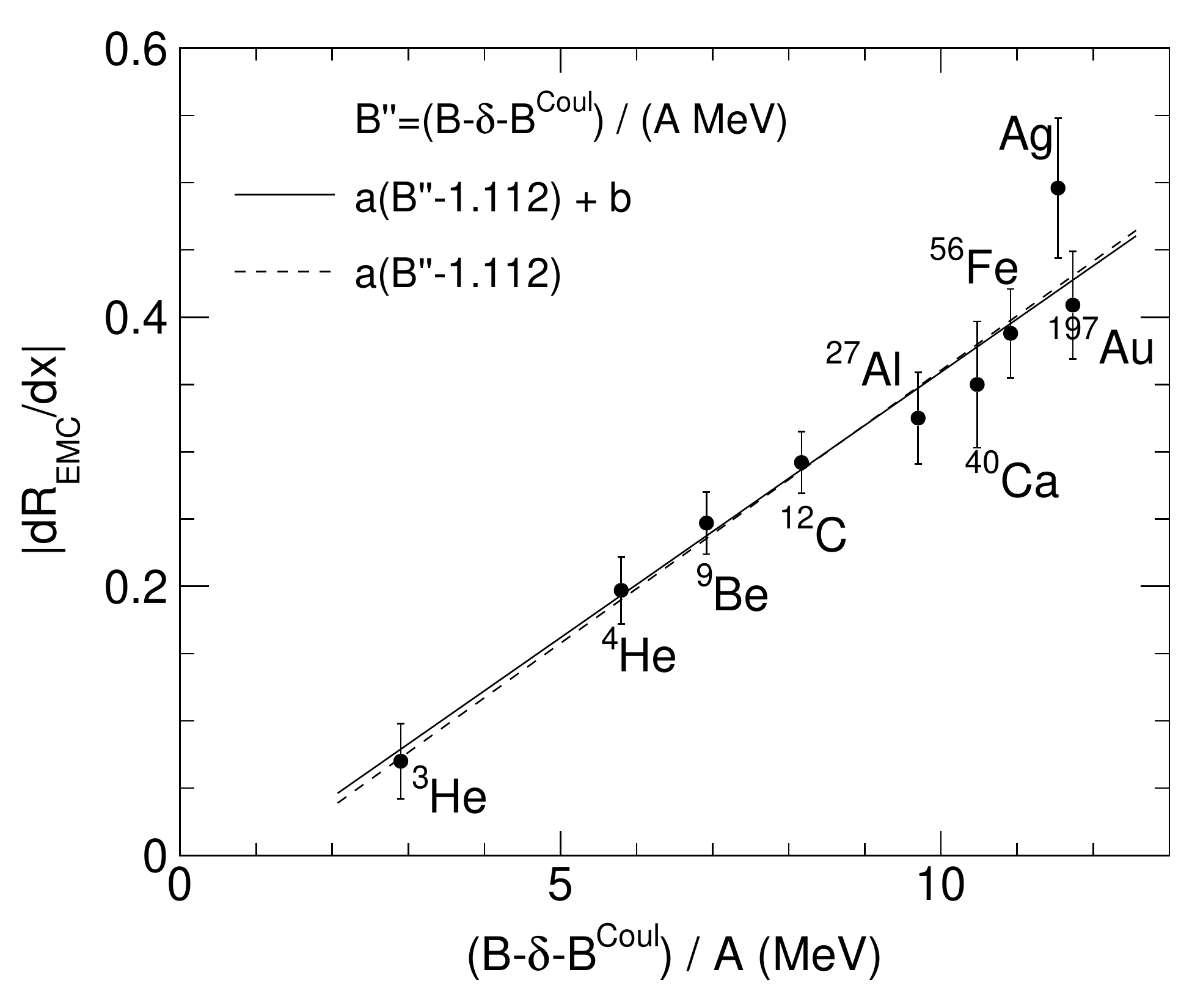}
\caption{
EMC slope $|dR_{\text{EMC}}/dx|$ as a function of per-nucleon ($B-\delta-B^{\text{Coul}}$).
The EMC slope data are taken from Ref. \cite{NuclDepenRev}.
For a silver target, both $^{107}$Ag and $^{109}$Ag have big natural abundances.
The weighted mean binding energy and weighted modified binding energy
are used for the Ag target.
}
\label{fig:EMC_ModB}
\end{center}
\end{figure}

As there is strong connection found between N-N SRC and the
EMC effect\cite{NuclDepenRev,src-emc1,src-emc2,src-emc3,src-emc4},
it is necessary to investigate the relationship between
the EMC effect and the modified nuclear binding with pairing energy
removed. The strength of the EMC effect is usually described by
the EMC slope $|dR_{\text{EMC}}/dx|$\cite{Seely2009,NuclDepenRev}.
A recent work \cite{Wang2015} has shown a good linear correlation
between the EMC effect and residual strong interaction energy (nuclear
binding with Coulomb contribution removed). Hence, it is interesting
and important to find out whether the linear correlation would be improved
by subtracting the pairing energy from the residual strong interaction energy.
Fig. \ref{fig:EMC_ModB} shows the correlation between EMC slopes and
($B-\delta-B^{\text{Coul}}$) per nucleon, in which $B^{\text{Coul}}$
is the contribution of Coulomb interaction to nuclear binding.
Amazingly, an excellent correlation is shown between these two quantities.
A linear fit is performed to the correlation, with both slope and intercept
parameters set free. $\chi^2/N_{df}$ of the fit is obtained to be 3.538/7=0.505.
The same linear fit to the correlation between $|dR_{\text{EMC}}/dx|$ and
$(B-B^{\text{Coul}})/A$ gives $\chi^2/N_{df}=7.050/7=1.01$.
Based on fit quality, a linear correlation between EMC slope and
$(B-\delta-B^{\text{Coul}})/A$ is noticeably better
than the linear correlation between EMC slope and $(B-B^{\text{Coul}})/A$.
Therefore the nuclear pairing effect does not play important role in
either N-N SRC or the EMC effect. Finally, we use the functional form
$a[(B-\delta-B^{\text{Coul}})/(A~\text{MeV})-1.112]$ to fit the linear correlation
between EMC slope and $(B-\delta-B^{\text{Coul}})/A$. The fit quality
is also good with $\chi^2/N_{df}=3.633/8=0.454$, though we do not
have an accurate calculation for the modified binding energy
$(B-\delta-B^{\text{Coul}})$ of the deuteron.

\section*{Acknowledgments}
One of us (Rong Wang) is very grateful to Wim Cosyn for the stimulating discussions
and providing us with calculations of number of N-N SRC. We thank Jarah Evslin
for reading the manuscript and the help of improving the English of the paper.
This work was supported by the National Basic Research Program (973 Program Grant No. 2014CB845406),
and Century Program of Chinese Academy of Sciences Y101020BR0.

\end{document}